\documentclass[a4paper,11pt]{article}
\pdfoutput=1 

\usepackage{jheppub} 

\usepackage[T1]{fontenc} 

\usepackage[compat=1.0.0]{tikz-feynman}
\usepackage{physics}
\usepackage{MnSymbol}
\usepackage{wasysym}
\usepackage{multirow}
\usepackage{comment}

\makeatletter
\newcommand*{\rom}[1]{\expandafter\@slowromancap\romannumeral #1@}
\makeatother

\begin{document}

\title{\boldmath {Thermodynamics of the Glueball Resonance Gas}}


\author[a]{Enrico~Trotti,}
\author[a]{Shahriyar~Jafarzade,}
\author[a,b]{Francesco Giacosa }


\affiliation[a]{Institute of Physics, Jan Kochanowski University,\\
ul. Uniwersytecka 7, 25-406, Kielce, Poland}
\affiliation[b]{Institute for Theoretical Physics, J. W. Goethe University,\\ Max-von-Laue-Str. 1, 60438 Frankfurt, Germany}

\emailAdd{trottienrico@gmail.com}
\emailAdd{shahriyar.jzade@gmail.com}
\emailAdd{francesco.giacosa@gmail.com
}

\abstract{We study the thermodynamic properties -pressure, entropy and trace anomaly- of a gas of glueballs that includes the glueball states obtained by various lattice simulations. We show that this model, called Glueball Resonance Gas (GRG) approach, describes well the thermal properties of the Yang-Mills sector of QCD below the critical temperature $T_c$, provided that $T_c$ is properly matched to the corresponding determination of the glueball masses, obtaining $T_c \sim 320 \pm 20$ MeV. The inclusion into the GRG of heavier glueballs not yet seen on the lattice, assuming that glueballs follow Regge trajectories as quark-antiquark states do, leads only to a small correction. We consider the contribution to the pressure of the interactions between scalar-scalar and tensor-tensor glueballs, which turn out to be also negligible.}

\maketitle
\section{Introduction}
The phase diagram of Quantum Chromodynamics (QCD) is one of the most relevant topics in high energy physics \cite{Ratti,Rischke:2003mt,Brambilla:2014jmp}. 
At nonzero temperature and zero quark chemical potential, QCD can be mainly described by a (relatively) weakly interacting hadron gas in the low
temperature regime and by the perturbative quark-gluon plasma (QGP) in the high temperature regime with a smooth cross-over phase transition between them.  

A plethora of approaches has been developed to study QCD at finite temperature, that range from models based on hadronic d.o.f. to those with quark d.o.f. or mixture of them \cite{Hansen:2006ee,Kovacs:2016juc,Sasaki:2012ybg,Heinz:2008cv,Heinz:2011xq}. Also various techniques, such as the CJT (Cornwall, Jackiw, Tomboulis) formalism \cite{Pilaftsis:2013xna}, the functional renormalization group (FRG) approach \cite{Alkofer:2018guy,Fu:2022gou,Pawlowski:2005xe,Koenigstein:2021syz}, and the S-matrix phase-shift formalism \cite{Broniowski:2015oha,Lo:2017sde,Lo:2017ldt,Lo:2017lym,Lo:2019who,Samanta:2020pez,Samanta:2021vgt} have been developed.
For our purposes, we mention in particular two well-defined methods: (i) the Hadron Resonance Gas (HRG) model, in which the thermodynamics of the confined phase of QCD is described (in its easiest formulation) by a gas of non-interacting hadrons \cite{Andronic:2008gu,Andersen:2011ug,Andronic:2012dm,Andronic:2021erx,Alba:2014eba,Torrieri:2004zz}; (ii) lattice QCD, which discretizes QCD on a 4D Euclidean grid. Lattice simulations are particularly successful to describe the physics of the phase transition, see e.g. Refs. \cite{Borsanyi:2010cj,Borsanyi:2012ve,Borsanyi:2013bia,Karsch:2003jg,Miller:2006hr,Sharma:2013hsa,HotQCD:2014kol}.

An important (and non-trivial) part of QCD is the Yang-Mills (YM) sector, in which only the gluonic part of the QCD Lagrangian is considered. Since the main features of QCD come from its non-abelian nature, YM retains both confinement and asymptotic freedom.  One of the major motivations to consider the YM part of QCD is the fact that in the limit of infinitely heavy quark masses, QCD is a pure gauge theory with an exact order parameter.
In this case the phase transition, which is expected to be first order, happens between the gluonic bound states at low temperature (the so-called glueballs, see below) and a gas of gluons at high temperature \cite{Levai:1997yx,Peshier:1995ty,Giacosa:2010vz}.
Examples of theoretical models and approaches devoted to the thermodynamics of pure YM are the T-Matrix approach \cite{Lacroix:2012pt}, the Polyakov-loop potential \cite{PhysRevD.29.1222}, functional methods \cite{Kondo:2015noa}, holographic QCD \cite{Zhang:2021itx}, quasi-particle models \cite{Mykhaylova:2019wci,Mykhaylova:2020pfk}, bag model \cite{Giacosa:2009,Pisarski:2006yk}, hard-thermal-loop perturbation theory \cite{Andersen:2011ug}, etc.
The YM  theory at nonzero $T$ has been also investigated on the lattice since long time,
one of the earliest lattice simulations describing a phase transition in the pure YM sector of QCD appeared more than 25 years ago \cite{Boyd:1996bx}.  Later on, various lattice works explored the thermal properties of QCD, e.g. \cite{Panero:2009tv,Lucini:2012gg,Borsanyi:2012ve,Caselle:2011fy,Caselle:2015tza,Alba:2016fku}.

The glueballs are the relevant low-energy hadronic degrees of freedom. Despite lacking a final experimental confirmation (even though some candidates do exist), they have been studied using several methods, e.g. in LQCD, where their spectrum is evaluated 
\cite{Chen:2005,Meyer:2004gx,Athenodorou:2020ani,Chen:2004bw}, as well as in various models \cite{Ochs:2013,Mathieu:2008,Amsler:1995td,Janowski:2014,Giac_ModellingGlue}. 
The existence of glueballs is reinforced by studies on pomeron and odderon trajectories \cite{Szanyi:2019kkn}.
One may therefore consider a glueball resonance gas (GRG) model as the analogous counterpart of the HRG model mentioned above (indeed, the GRG is nothing else than a subset of the HRG).
In Ref. \cite{Buisseret:2010mop}, it was shown that a GRG with the lattice available glueballs of Ref. \cite{Chen:2005} describes quite well the low-temperature part of the YM pressure, but is somewhat too low when approaching $T_c$. Because of that, the inclusion of an additional Hagedorn contribution (with density of states $\rho(m) \propto e^{m/T_H}$ \cite{Hagedorn:1965st}) was proposed as a possible explanation of the lattice thermodynamic simulations close to the critical temperature $T_c$ \cite{Meyer:2009tq}. 

In this work, we revisit the study of the thermodynamics of YM in the following way: we build the GRG pressure by using the lattice results for the glueball mass spectrum obtained in the three works of Refs. \cite{Chen:2005,Meyer:2004gx,Athenodorou:2020ani} and compare the outcomes to the corresponding lattice thermodynamic results of Ref. \cite{Borsanyi:2012ve}. For each mass compilation of Refs. \cite{Chen:2005,Meyer:2004gx,Athenodorou:2020ani}, we deduce the corresponding value of the critical temperature $T_c$ accordingly. Namely, the lattice results of Ref. \cite{Borsanyi:2012ve} are expressed as function of $T/T_c$, thus a comparison is possible only after such a link is built.  
In particular, we shall show that the GRG pressure with the masses of Ref. \cite{Athenodorou:2020ani} agrees quite well with the lattice results almost up to corresponding $T_c \sim 320 \pm 20$ MeV without the need of additional contributions. The critical temperature turns out to be larger than the commonly used value of $T_c \sim 260$ MeV. Interestingly, a value of about $300$ MeV for pure YM was obtained within functional methods in Ref. \cite{Braun:2010cy}.

Furthermore, we also investigate further contributions to the GRG results along two directions. 
(i) We include  the effect of heavier glueballs which were not yet seen on lattice simulations, but are nevertheless expected to exist. In practice, for each $J^{PC}$ we add 10 radially excited states from Regge trajectories in order to see the increment in pressure and trace anomaly. (ii) By applying the S-matrix formalism, we study the contributions to the pressure that arise from the interactions among scalar-scalar and tensor-tensor glueballs, which are the lowest lying states in the glueball spectrum.  We find that both effects are very small and in practice negligible. Thus, the GRG built with non-interacting glueballs seen on the recent lattice simulation of Ref. \cite{Athenodorou:2020ani} seems to provide a rather good approximation of YM thermodynamics.

The article is built as it follows: in Sec. \ref{sec:grg} we introduce the glueball resonance gas in general and its implementation by using the masses from lattice QCD. The contributions to the thermodynamic quantities that arise from further excited states and from glueball interactions are presented in Sec.  \ref{corr}. Conclusion can be found in Sec. \ref{concl}. Further elaboration on the glueball scattering is recalled in App. \ref{app:TT}.


\section{Thermodynamics of the Glueball Resonance Gas}
\label{sec:grg}

The basic quantities to describe thermodynamic properties of the GRG are the pressure $p_i$ and the energy density $\epsilon_i$ of the $i$-$th$ glueball which,
for non-interacting particles, reads (e.g. Ref. \cite{Ratti}):
\begin{equation}
    p_{i}=-(2J_i+1)T\int_{0}^\infty \dfrac{k^2}{2 \pi^2} \ln{\Big(1-e^{-\frac{\sqrt{k^2+m_i^2}}{T}}\Big)}dk \text{ ,}
    \label{eq.pfree}
\end{equation}
and 
\begin{equation}
\epsilon_i =(2J_i+1)\int_0^{\infty} \dfrac{k^2}{2 \pi^2} \dfrac{\sqrt{k^2 +m_i^2}}{\exp \Biggl[ \dfrac{\sqrt{k^2 +m_i^2}}{T}  \Biggr] - 1} dk \text{ ,}
\end{equation}
where $J_i$ is the total spin of the $i$-th state.
By considering $N$ glueballs, the total pressure and energy density of the GRG are given by:
 \begin{equation}
    p^{GRG} \equiv p = \sum_{i=1}^{N} p_i \text{  ,  } \epsilon^{GRG} \equiv \epsilon = \sum_{i=1}^{N} \epsilon_k \text{ .}
\end{equation}
Other two important quantities are the trace anomaly $I$ and the entropy density $s$ defined as:
\begin{equation}
    I=\epsilon-3p\,,\quad s = \frac{p+\epsilon}{T} \text{ .}
    \label{IS}
\end{equation}
It is also convenient to introduce the dimensionless pressure $\hat{p}=p/T^4$ and energy density $\hat{\epsilon}=\epsilon/T^4$, as well as the dimensionless trace anomaly $\hat{I} = \hat{\epsilon}-3\hat{p}$ and entropy $\hat{s}=s/T^3$. 

The equations above allow us to calculate the GRG pressure and energy density by using the masses from the lattice works listed in Tables \ref{tab:lattparam} and \ref{tablemass} and compare them to the  corresponding lattice predictions of Ref \cite{Borsanyi:2012ve}. 
In the latter work, the thermodynamic quantities are expressed as function of the ratio $T/T_c$.
Thus, for a meaningful comparison of the GRG results obtained by using different lattice results for the glueball masses, it is of primary importance to consider the location of the critical temperature separately for each given lattice simulation. 
In order to achieve that, we recall that the critical temperature $T_c$ is related to the so called Lambda parameter $\Lambda_{MS}$ and to the QCD string tension $\sigma$ through the following relations \cite{Gockeler:2005rv,Guagnelli:1998ud,Boyd:1996bx}: 
\begin{align}
 T_c&=1.26(7) \cdot \Lambda_{MS} =  1.26(7) \cdot 0.614(2) \cdot r_0^{-1}  \text{ ,}\label{eq:TC1}
 \\
 T_c&=0.629(3) \cdot \sqrt{\sigma} \text{ .}
 \label{eq:TC2}
\end{align}
In Table \ref{tab:lattparam} we summarize the lattice parameters for the simulations of the glueball mass spectrum and we thus deduce the corresponding critical temperature (together with an estimate of its uncertainty) upon using the equations above. 
Note, the glueball masses in \cite{Athenodorou:2020ani} displace only statistical errors.

The use of a different scale $r_0=0.472(5)$ fm (equivalent to 
$ r_0^{-1}=418(5)$ MeV) in Ref. \cite{Athenodorou:2020ani}, motivated by \cite{Sommer:2014mea}, leads to slightly different central values for the glueball masses w.r.t. Ref. \cite{Chen:2005}, but they are still comparable within errors. Instead, both $T_c$ and the masses of Ref. \cite{Meyer:2004gx} are lower.

The masses of the glueballs with quantum numbers $J^{PC}=0^{++},2^{++},0^{-+},2^{-+},1^{+-}$ are also obtained by using a different technique in Ref. \cite{Bonanno:2022yjr} for $N_c =6$: quite remarkably, the  outcomes agree with a $2-5\%$ accuracy with the results of Ref. \cite{Athenodorou:2020ani}, confirming the validity of both the large-$N_c$ limit and the masses in \cite{Athenodorou:2020ani}. 
It is also interesting to point out that recent progress in functional methods and Bethe-Salpeter approaches leads to a spectrum which is comparable to that obtained by lattice simulations \cite{Huber:2021yfy, Sanchis-Alepuz:2015,Huber:2020ngt}.





\begin{table}[ht]
    \centering
    \begin{tabular}{|c|c|c|c|}
    \hline
        LQCD papers & Number of  & Lattice Parameter & $T_c$  (using\\
     & glueballs &  & Eqs. (\ref{eq:TC1})-(\ref{eq:TC2}))\\
        \hline
          Chen et.al \cite{Chen:2005} & $12$ & $r_0^{-1}=410(20)$ \text{MeV}  & $317\pm 23$ \text{MeV}\\\hline
        Meyer \cite{Meyer:2004gx} & 22 & $\sqrt{\sigma}=440(20)$ \text{MeV} & $277\pm 13$ \text{MeV} \\\hline
        Athenodorou and Teper \cite{Athenodorou:2020ani} & 20 & $r_0^{-1}=418(5)$ MeV& $323\pm 18 $ \text{MeV} \\\hline
    \end{tabular}
    \caption{Parameters entering lattice simulations of the glueball mass spectrum.}
    \label{tab:lattparam}
\end{table}

\begin{table}[ht]
\scriptsize
  \begin{center}
  
 \begin{tabular}{|c|c|c|c|c|c|c|c|}
    \hline
    \rule{0pt}{2em} $n\,J^{PC}$ & \multicolumn{3}{|c|}{M[MeV]}& \rule{0pt}{2em} $n\,J^{PC}$ & \multicolumn{3}{|c|}{M[MeV]}  \\\hline
     & Chen et.al. \cite{Chen:2005} & Meyer \cite{Meyer:2004gx} &A \& T \cite{Athenodorou:2020ani}  &   & Chen et.al. \cite{Chen:2005}& Meyer \cite{Meyer:2004gx} & A \& T \cite{Athenodorou:2020ani}  \\\hline
     $\textbf{1\, 0}^{++}$ & 1710(50)(80)  & 1475(30)(65)& 1653(26) & $\textbf{1\,1}^{--}$ & 3830(40)(190)  & 3240(330)(150)& 4030(70)
 \\\hline
     $\textbf{2\,0}^{++}$ &   & 2755(30)(120) & 2842(40) & $\textbf{1\,2}^{--}$ & 
 4010(45)(200)  &3660(130)(170)& 3920(90)
   \\\hline
      $\textbf{3\,0}^{++}$ &    & 3370(100)(150) & & $\textbf{2\,2}^{--}$  & &3740(200)(170)&
 \\\hline
     $\textbf{4\,0}^{++}$ &  
 & 3990(210)(180) &  & $\textbf{1\,3}^{--}$& 4200(45)(200)  &4330(260)(200)& 
   \\\hline
    $\textbf{1\,2}^{++}$ & 2390(30)(120)    & 2150(30)(100) & 2376(32) & $\textbf{1\,0}^{+-}$ & 4780(60)(230) & &
 \\\hline
    $\textbf{2\,2}^{++}$ &   & 2880(100)(130)  &  3300(50) & $\textbf{1\,1}^{+-}$ & 2980(30)(140)   & 2670(65)(120)& 2944(42)
 \\\hline
    $\textbf{1\,3}^{++}$ & 3670(50)(180)  & 3385(90)(150)& 3740(70)
 & $\textbf{2\,1}^{+-}$ &  & 
 &3800(60)
 \\\hline
    $\textbf{1 4}^{++}$ &    & 3640(90)(160) & 3690(80) & $\textbf{1\,2}^{+-}$ & 4230(50)(200) &  &4240(80)  \\\hline
    $\textbf{1\,6}^{++}$ &  &   4360(260)(200) &  & $\textbf{1\,3}^{+-}$ & 3600(40)(170)  & 3270(90)(150) & 3530(80)
 \\\hline
     $\textbf{1\,0}^{-+}$ & 2560(35)(120)  & 2250(60)(100)   & 2561(40)& $\textbf{2\,3}^{+-}$ & & 3630(140)(160) & 

    \\\hline
      $\textbf{2\,0}^{-+}$ &     & 3370(150)(150) & 3540(80) & $\textbf{1\,4}^{+-}$ & & &4380(80)   \\\hline
       $\textbf{1\,2}^{-+}$ & 3040(40)(150)   &2780(50)(130)& 3070(60)
    & $\textbf{1\,5}^{+-}$& &   4110(170)(190)&
       \\\hline
        $\textbf{2\,2}^{-+}$ &   & 3480(140)(160) & 3970(70)  & & & &   \\\hline
         $\textbf{1\,5}^{-+}$ &  &  3942(160)(180)&   & & & &   \\\hline
          $\textbf{1\,1}^{-+}$ &  &  & 4120(80)   & & & &   \\\hline
          $\textbf{2\,1}^{-+}$ &  &  & 4160(80)   & & & &   \\\hline
          $\textbf{3\,1}^{-+}$ &  &  & 4200(90)   & & & &   \\\hline
        
    \end{tabular}  
    \caption{The values of the glueball masses as given in three lattice works.}
     \label{tablemass}
 \end{center}
 \end{table}
 \noindent

\begin{figure*}[h]
        \includegraphics[scale=0.38]{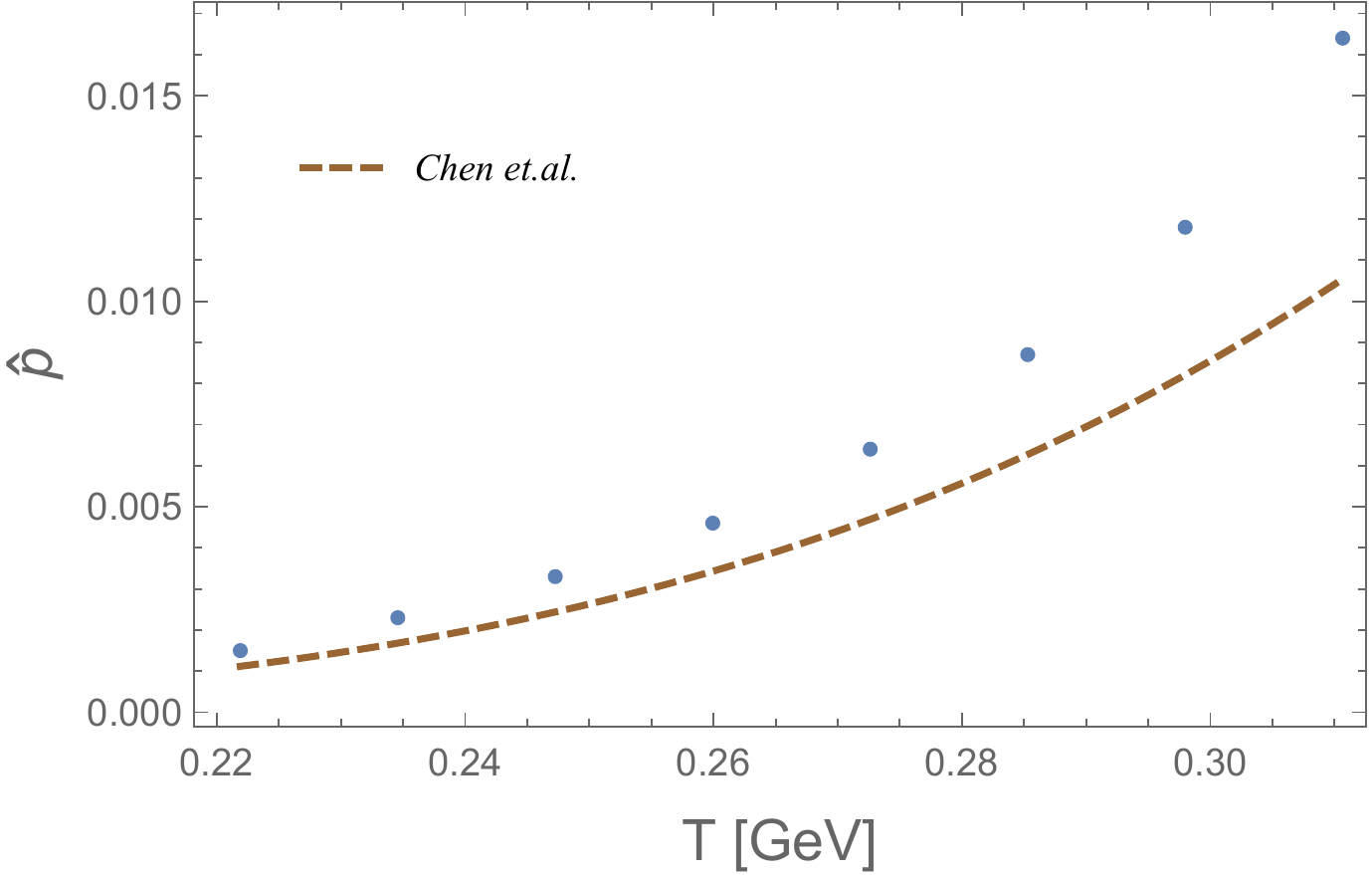}~\includegraphics[scale=0.38]{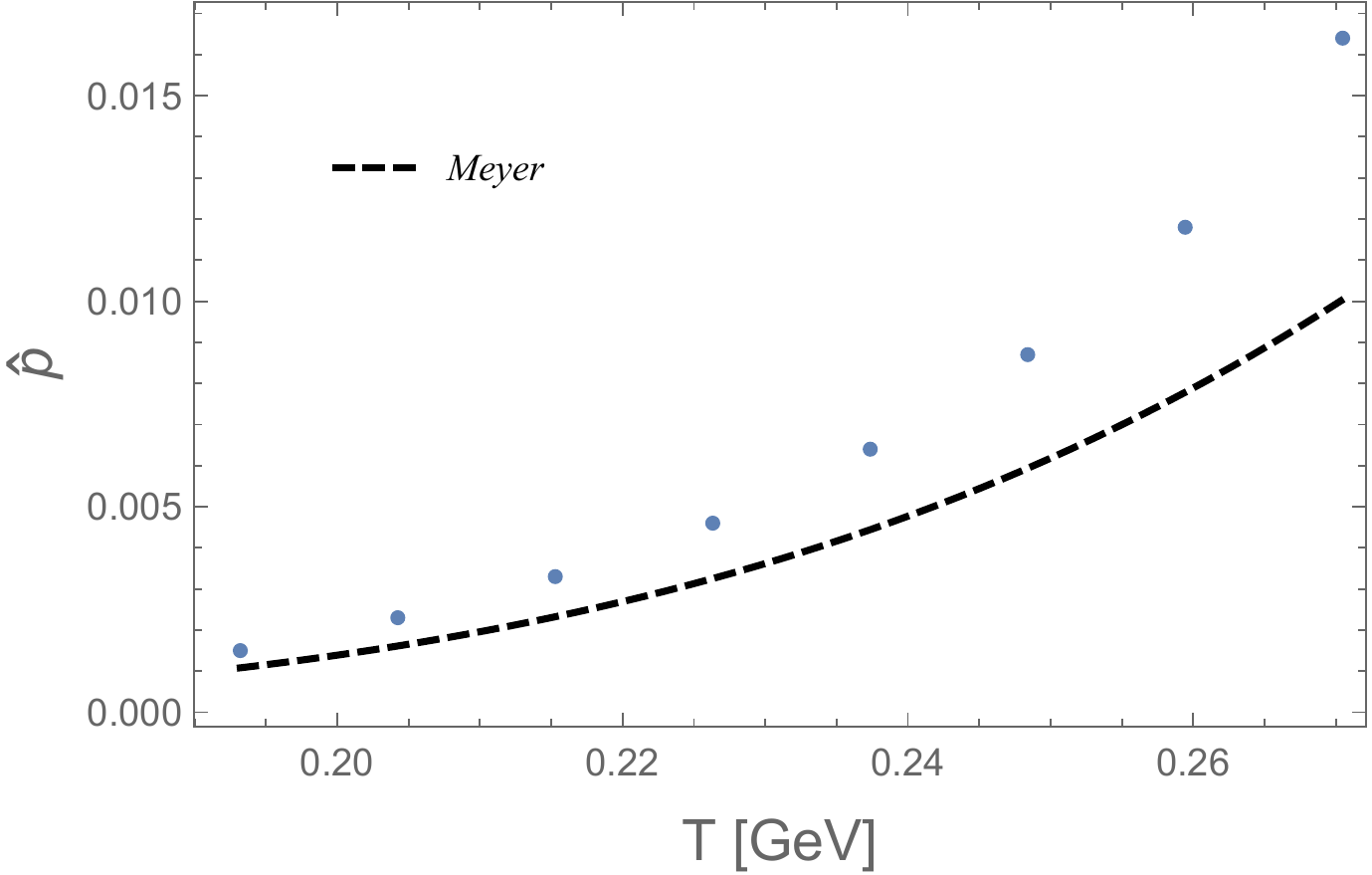}~\includegraphics[scale=0.38]{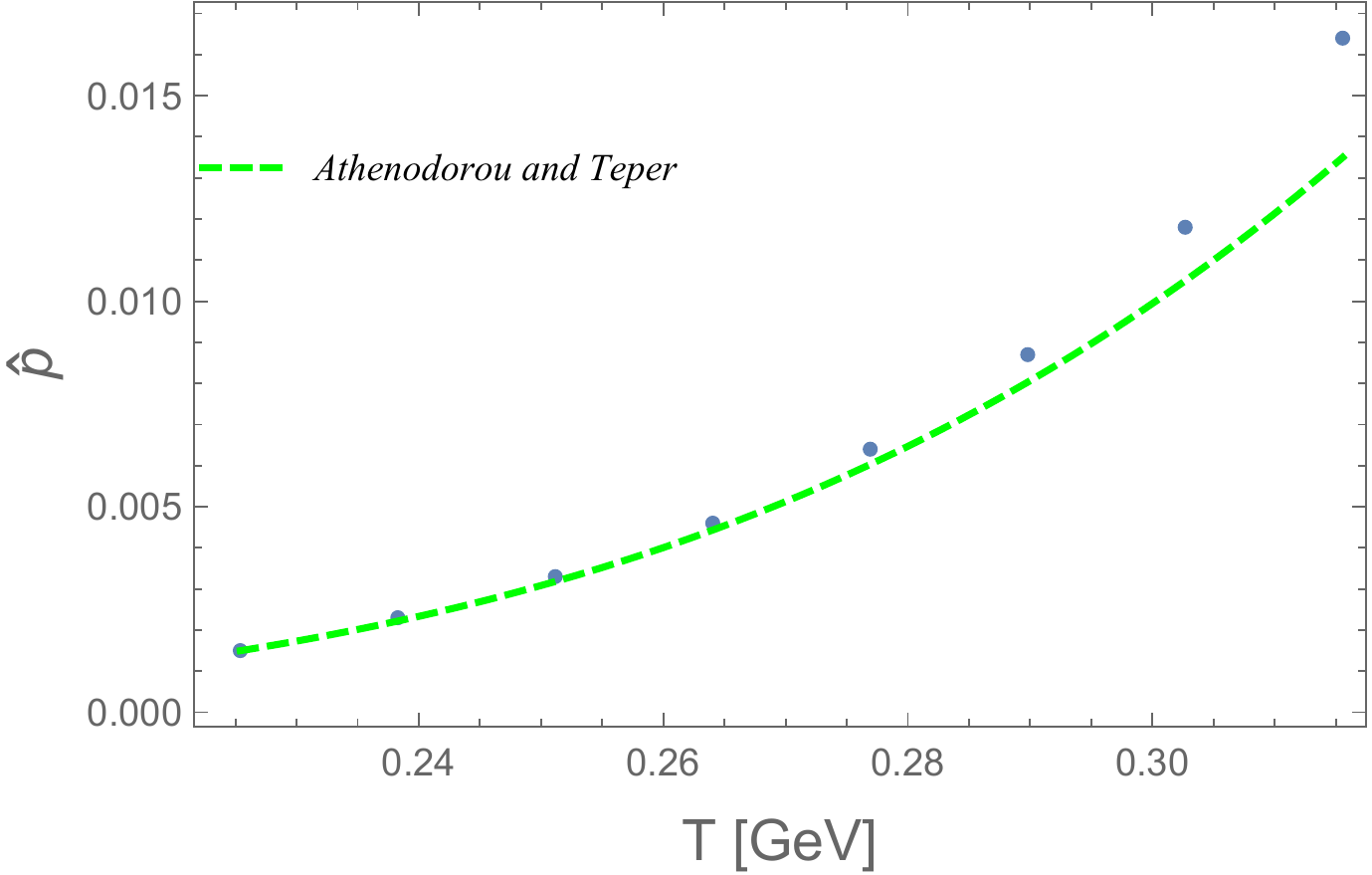}\\
        \caption{Normalized pressure of the GRG as function of the temperature for three different sets of lattice masses compared with the pressure evaluated in Ref.  \cite{Borsanyi:2012ve}. Left: GRG with masses from \cite{Chen:2005}; center: GRG with masses from \cite{Meyer:2004gx}; right: GRG with masses from \cite{Athenodorou:2020ani}.}
        \label{Pfree}
   \end{figure*}
\begin{figure*}[h]
        \centering
        \includegraphics[scale=0.38]{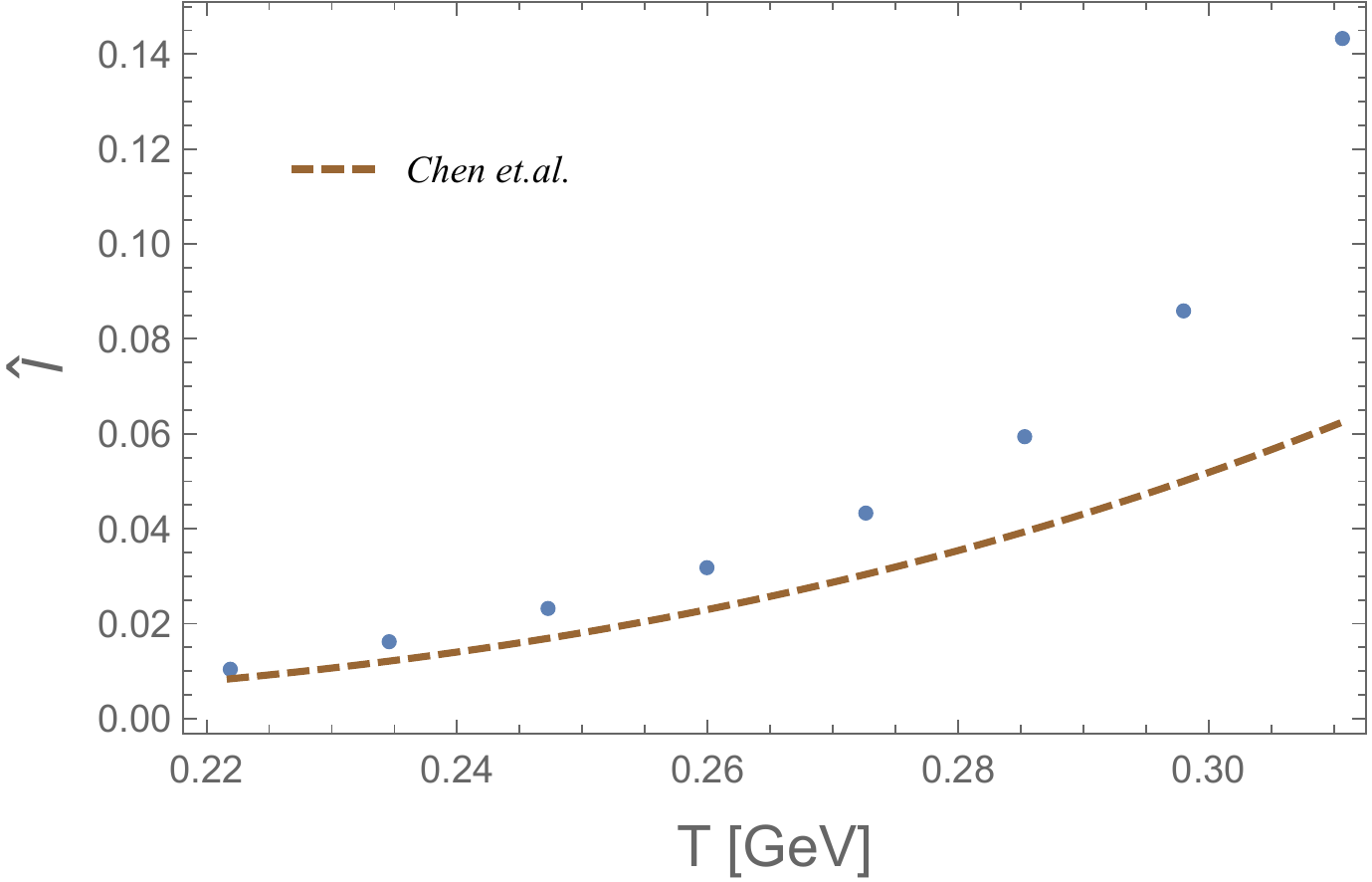}~\includegraphics[scale=0.38]{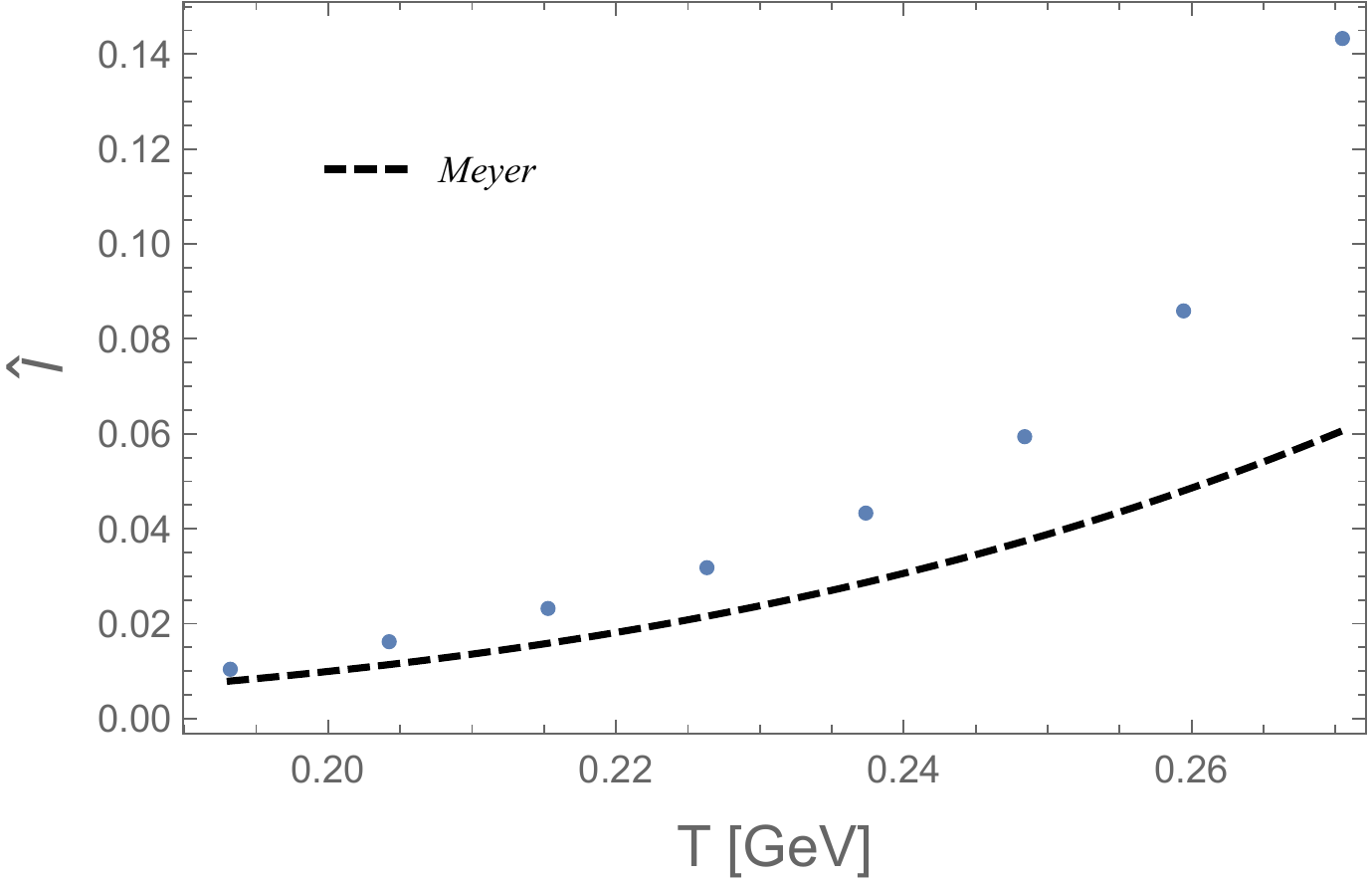}~\includegraphics[scale=0.38]{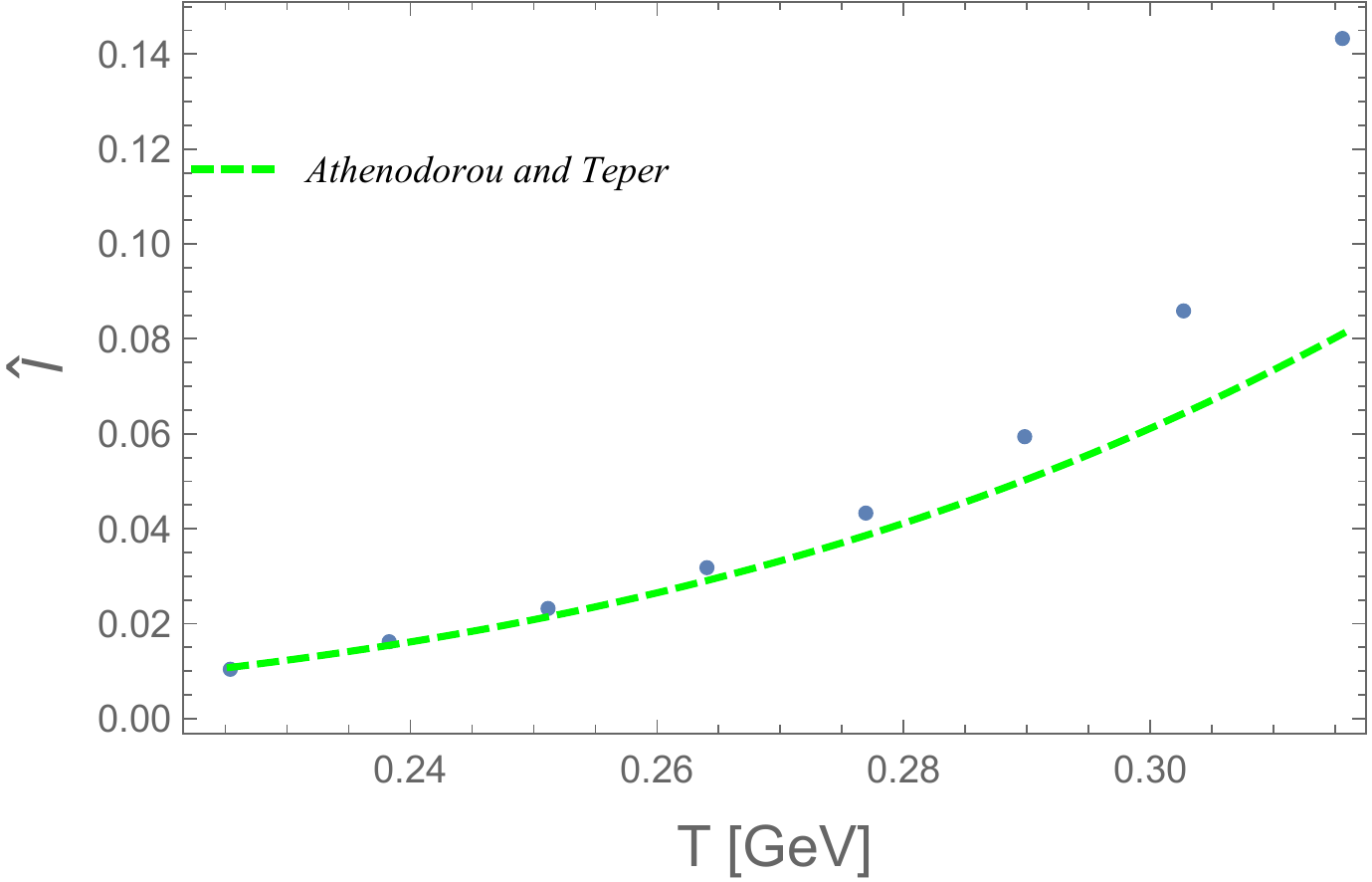}\\
        \caption{Normalized trace anomaly of the GRG as function of the temperature for three different sets of lattice masses compared with the values given in \cite{Borsanyi:2012ve}. Left: GRG with masses from  \cite{Chen:2005}; center: GRG with masses from \cite{Meyer:2004gx}; right: GRG with masses from \cite{Athenodorou:2020ani}.}
        \label{Afree}
   \end{figure*}
\begin{figure*}[h]
        \centering
 \includegraphics[scale=0.38]{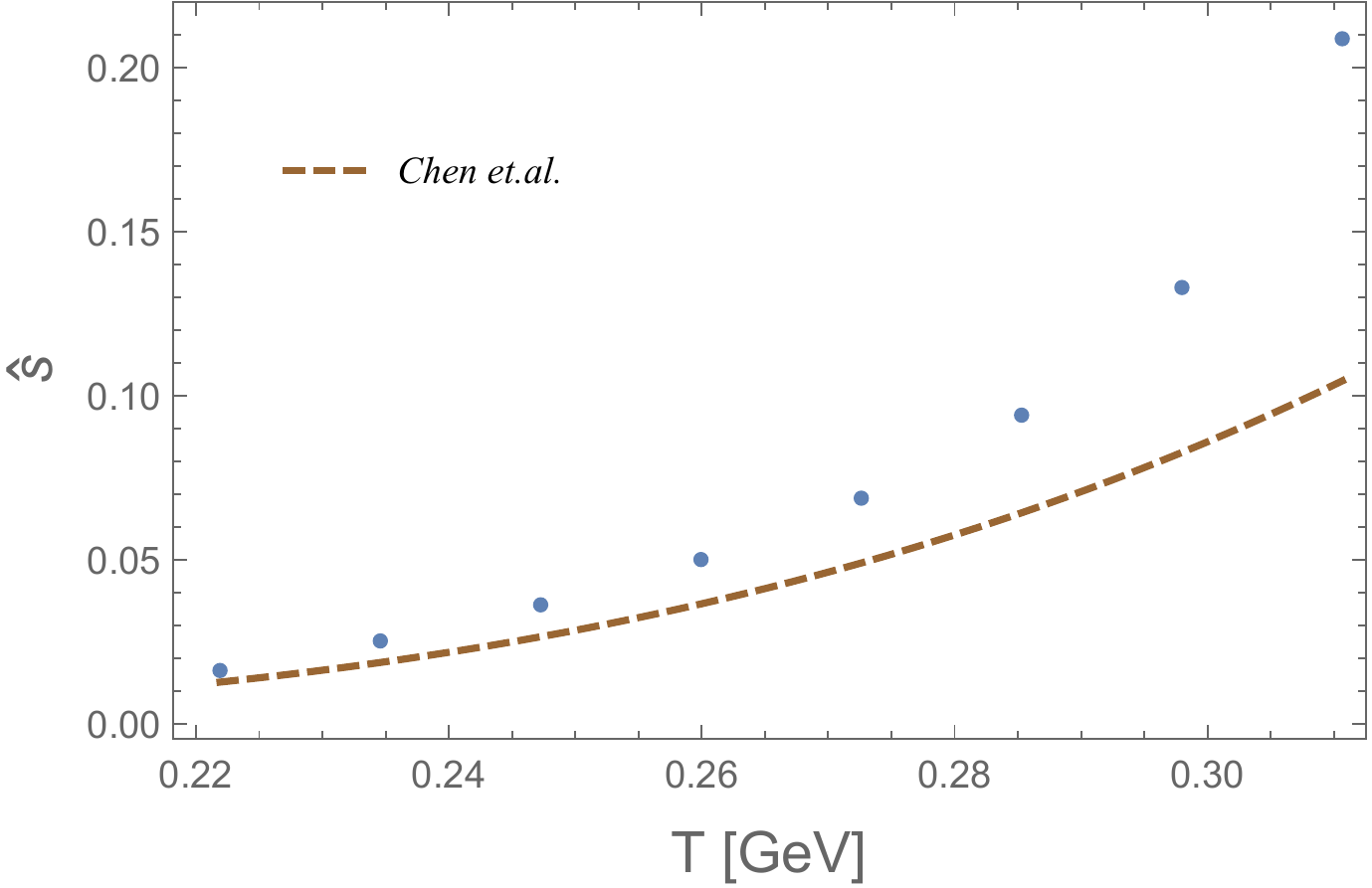}~\includegraphics[scale=0.38]{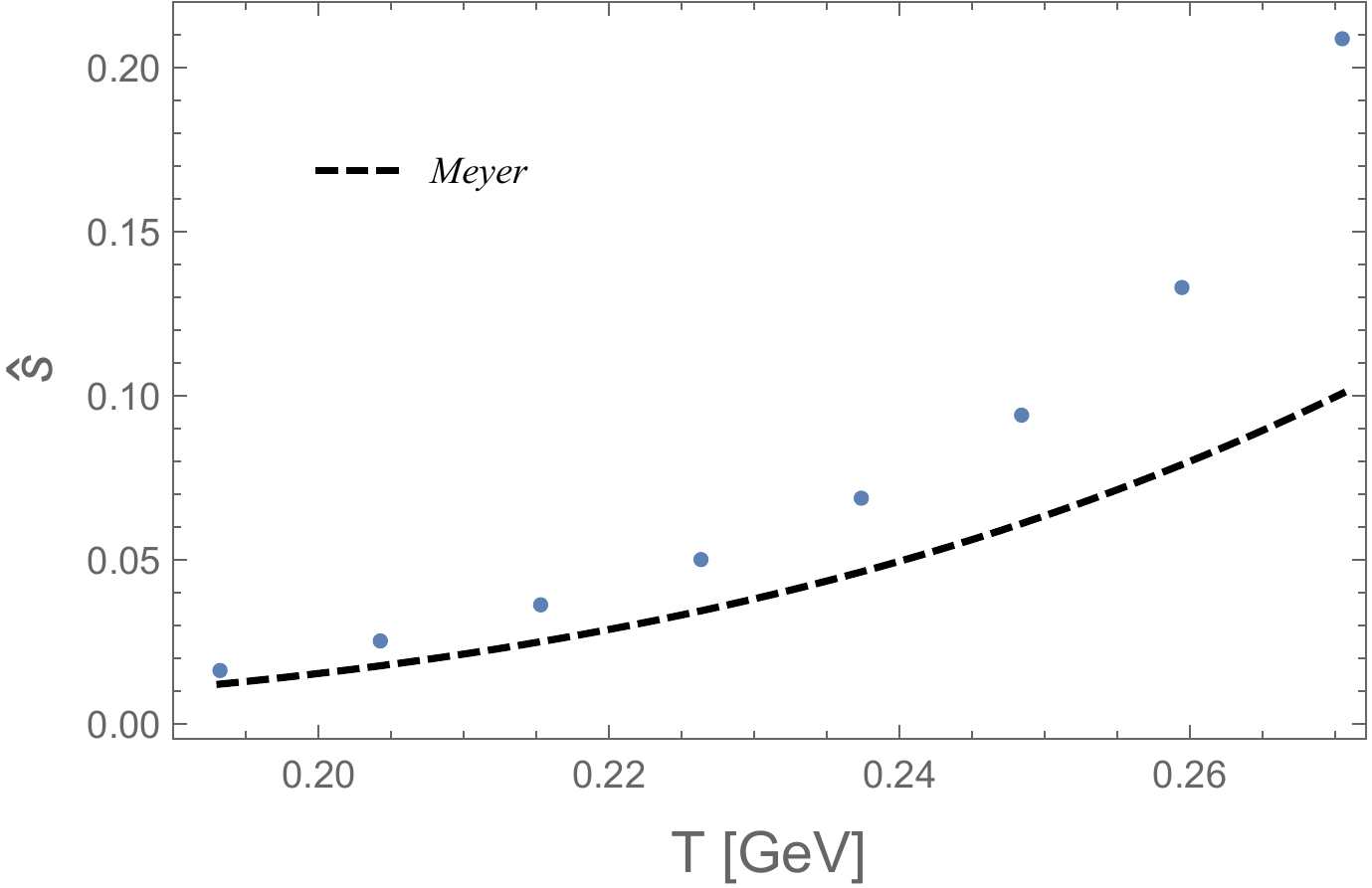}~\includegraphics[scale=0.38]{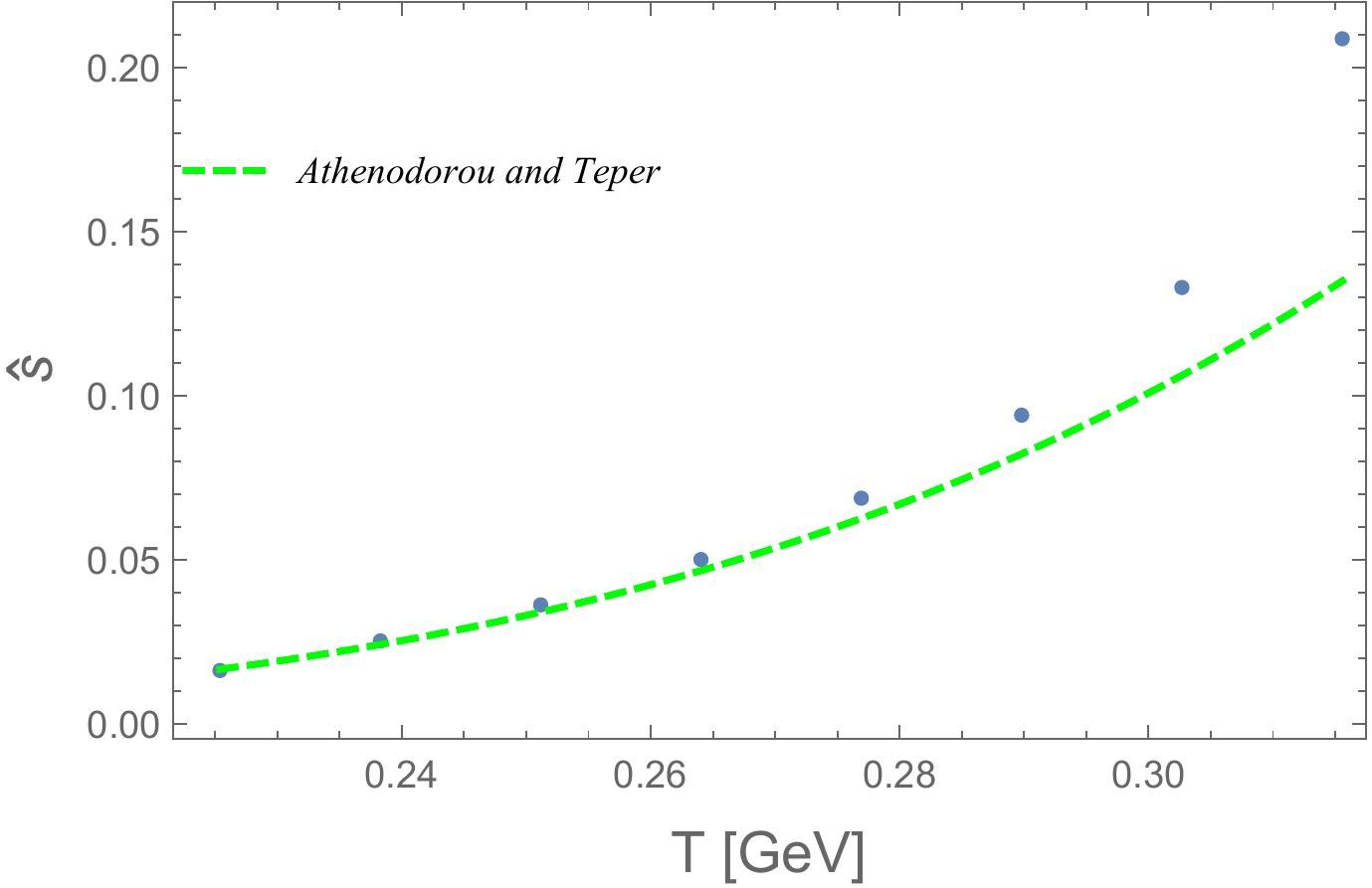}        \\
        \caption{Normalized entropy of the GRG as a function of the temperature for three different sets of lattice masses within GRG model, compared with the values given in \cite{Borsanyi:2012ve}. Left: GRG with masses from \cite{Chen:2005}; center: GRG with masses from \cite{Meyer:2004gx}; right: GRG with masses from \cite{Athenodorou:2020ani}.}
        \label{Sfree}
   \end{figure*}

In Fig. \ref{Pfree} we compare the results of the GRG for the parameters reported in Table \ref{tab:lattparam} to the thermodynamic lattice results of Ref. \cite{Borsanyi:2012ve}, in which the values of $T_c$ reported in Table \ref{tab:lattparam} have been used. Similar plots are obtained for the trace anomaly and the entropy density, see Fig. \ref{Afree} and Fig. \ref{Sfree}.
The GRG with the glueball spectrum from the most recent lattice work \cite{Athenodorou:2020ani} (right panel) approximates quite well the thermodynamic predictions from LQCD, especially for $T/T_c \leq 0.9$. 
As an interesting consequence, the critical temperature $T_c \approx 323 \pm 18$ MeV turns out to be somewhat larger than the usually employed reference value  $260$-$270$ MeV estimated in Ref. \cite{Borsanyi:2012ve}, but is in agreement with the FRG result of Ref. \cite{Braun:2010cy}.

Some considerations are in order:

(i) The GRG results evaluated via Refs. \cite{Chen:2005,Meyer:2004gx} (left and central panels of Figs. \ref{Pfree}-\ref{Sfree}) are well below the corresponding lattice curves. A possible explanation would be that some important contributions are missing, e.g. Ref. \cite{Borsanyi:2012ve}. Yet, as we shall see later on, this may not necessarily be the case.

(ii) The agreement of the plain non-interacting GRG evaluated with 20 lattice glueball states found in Ref. \cite{Athenodorou:2020ani} with the lattice calculation of Ref. \cite{Borsanyi:2012ve} for $T/T_c \leq 0.9$  is quite remarkable (right panels of Figs. \ref{Pfree}-\ref{Sfree}). Indirectly, these results show that the masses of Ref. \cite{Athenodorou:2020ani} are favoured.  In this case, eventual additional contributions to the GRG -if sizable- would rather spoil the agreement, but this scenario is not supported by the outcomes presented in the next Section.

(iii) The results discussed in  point (ii) are in agreement with the expectations of the large-$N_c$ limit of the YM theory \cite{tHooft:1974pnl,Witten:1979kh,Lebed:1998st}. Namely, the interaction among glueballs scales as $N_c^{-2}$, thus glueballs can be seen, in first approximation, as non-interacting bosons within the YM sector. In general, it is not clear if and for which quantities $N_c = 3$ can be considered as a large number, but the results of the right panels of Figs. \ref{Pfree} and \ref{Afree} seem to corroborate this view for what concerns YM thermodynamics in the confined phase.



\section{Corrections to the thermodynamic quantities}
\label{corr}
There are two main additional contributions to the GRG  that should be taken into account.  
\begin{itemize}
    \item Further excited states, that have not been observed in lattice simulations yet. Namely, an infinite tower of glueball states is expected to exist \cite{Witten:1979kh}.  We shall study their role by using Regge trajectories for the glueballs, see Sec. \ref{regge}.
    \item Interactions between glueballs. In fact, the GRG can be considered as a free gas only in first approximation, but interactions among glueballs and decays of heavier unstable glueballs take place. Here, we estimate in Sec. \ref{int} the role of the interaction for the lightest gluonic states (which are also the potentially most relevant ones): the scalar and the tensor glueballs.
\end{itemize}

\subsection{Contribution of heavier glueballs}
\label{regge}

As suggested in several works (e.g. Refs. \cite{Masjuan:2012gc,Anisovich:2000kxa}), light conventional quark-antiquark ($\Bar{q}q$) states can be systematically grouped into planes in a $(n,J,M^2)$-three dimensional space, where $M$ refers to the masses, $n$ to the radial quantum number, and $J$ to the total angular momentum.
These are the so called Regge trajectories, that (in principle) can be used to predict the masses of $\Bar{q}q$ states for an arbitrary value of $J$ and $n$:
\begin{equation}
    M^2(n,J^{PC})=a(n+J)+b_{J^{PC}}\,.
\label{eq:regge}
\end{equation}

Here, we apply the idea of the Regge trajectories to the YM sector. Despite the concept of Regge trajectories being analogous for both $\Bar{q}q$ states and glueballs, the applicability in the last case is complicated by the shortage of information about glueballs. In order to group the particles into different planes, many factors have to be considered (e.g. the number of gluons inside a certain $J^{PC}$ glueball) \cite{Meyer:2004gx}. Additionally, while Chen et. al \cite{Chen:2021bck} suggested that, within the relativistic framework, the ground-state glueballs with quantum numbers $J^{PC}=1^{++}$ and $1^{-+}$ do not exist as their corresponding currents vanish, other works \cite{Chen:2004bw} provide a mass for these glueballs. This ambiguity suggests that not all the possible $J^{PC}$ could in principle be found in lattice. 

At present, lattice masses are not enough to separate all the glueballs into different sectors and it is not possible to find a trajectory for each of them. Thus, there is a certain freedom in choosing how to group the glueballs into Regge plans. Here, we used the masses from Ref. \cite{Athenodorou:2020ani} and choose those states having ground state masses $\leq 3$ GeV.
Thus, we consider the five quantum numbers $J^{PC}=0^{++},2^{++},0^{-+},2^{-+},1^{+-}$ (for which the ground state is clearly observed in all three lattice simulations) and perform the following fit:
\begin{align}
   \chi^2(a,b_{0^{++}},b_{2^{++}},b_{0^{-+}},b_{2^{-+}},b_{1^{+-}})=\sum_{J^{PC}}
   \Biggl( \dfrac{M(n,J^{PC})-M^{\text{lat}}(n,J^{PC})}{\delta M^{\text{lat}}(n,J^{PC})} \Biggr)^2\,,
\label{chi2}
\end{align}
where $a$ and $b_{J^{PC}}$ are the parameters of the Regge trajectories, $M^{\text{lat}}$ are the lattice masses of the corresponding glueballs, and $\delta M^{\text{lat}}$ are the errors in the masses (see AT columns in Table \ref{tablemass}). The values of the parameters obtained in the fit, together with the masses used, are listed in Tab. \ref{tab:fit}.
By minimizing the $\chi^2$ we get $\chi^2_{\text{d.o.f.}}=0.84$ and $a=5.49 \pm 0.17\, \text{GeV}^2$ , showing that a unique slope is consistent with lattice data from \cite{Athenodorou:2020ani}: this is on its own an interesting outcome. 

\begin{figure*}[h]
        \centering
        \includegraphics[scale=0.5]{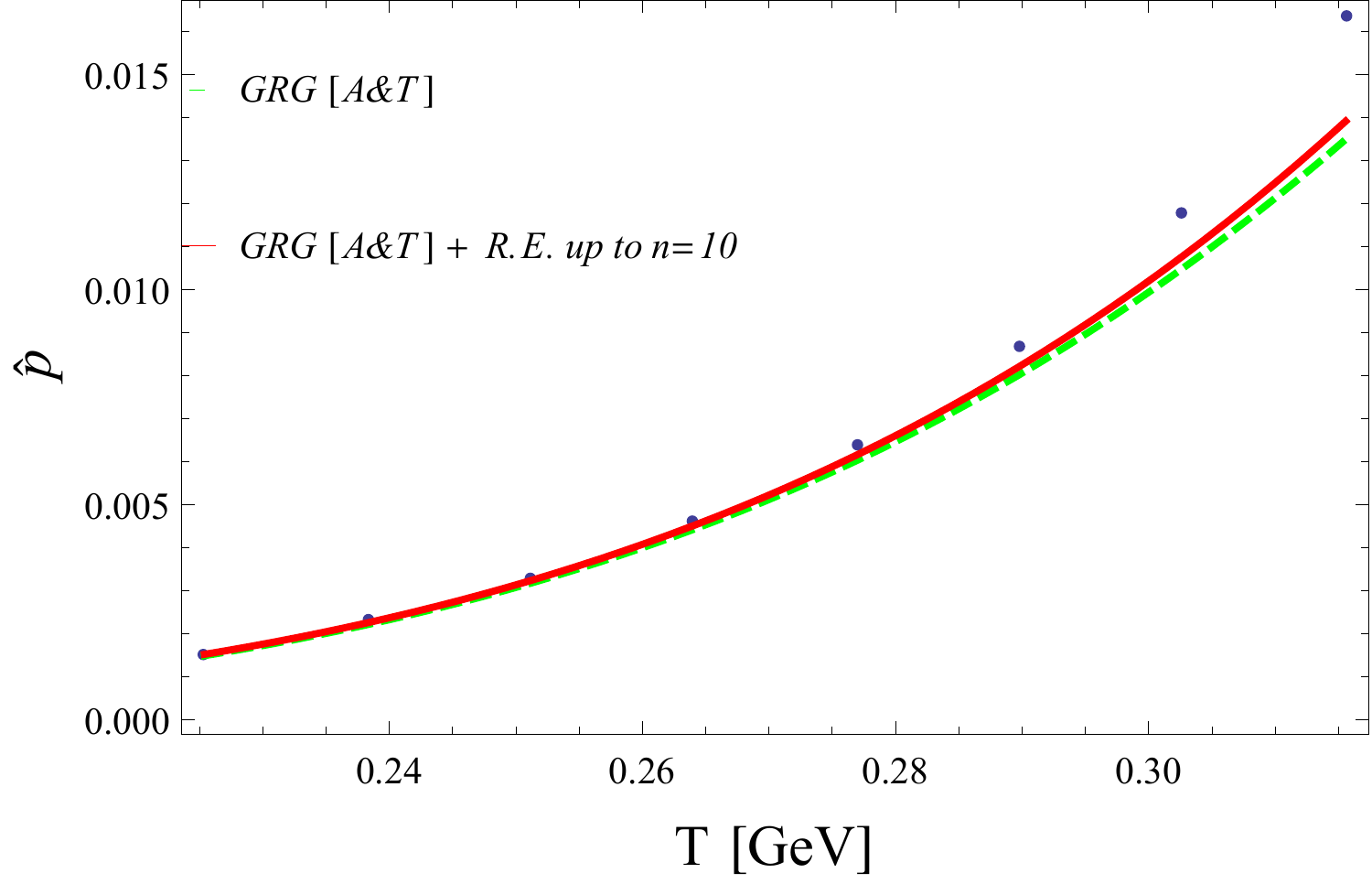}~\includegraphics[scale=0.5]{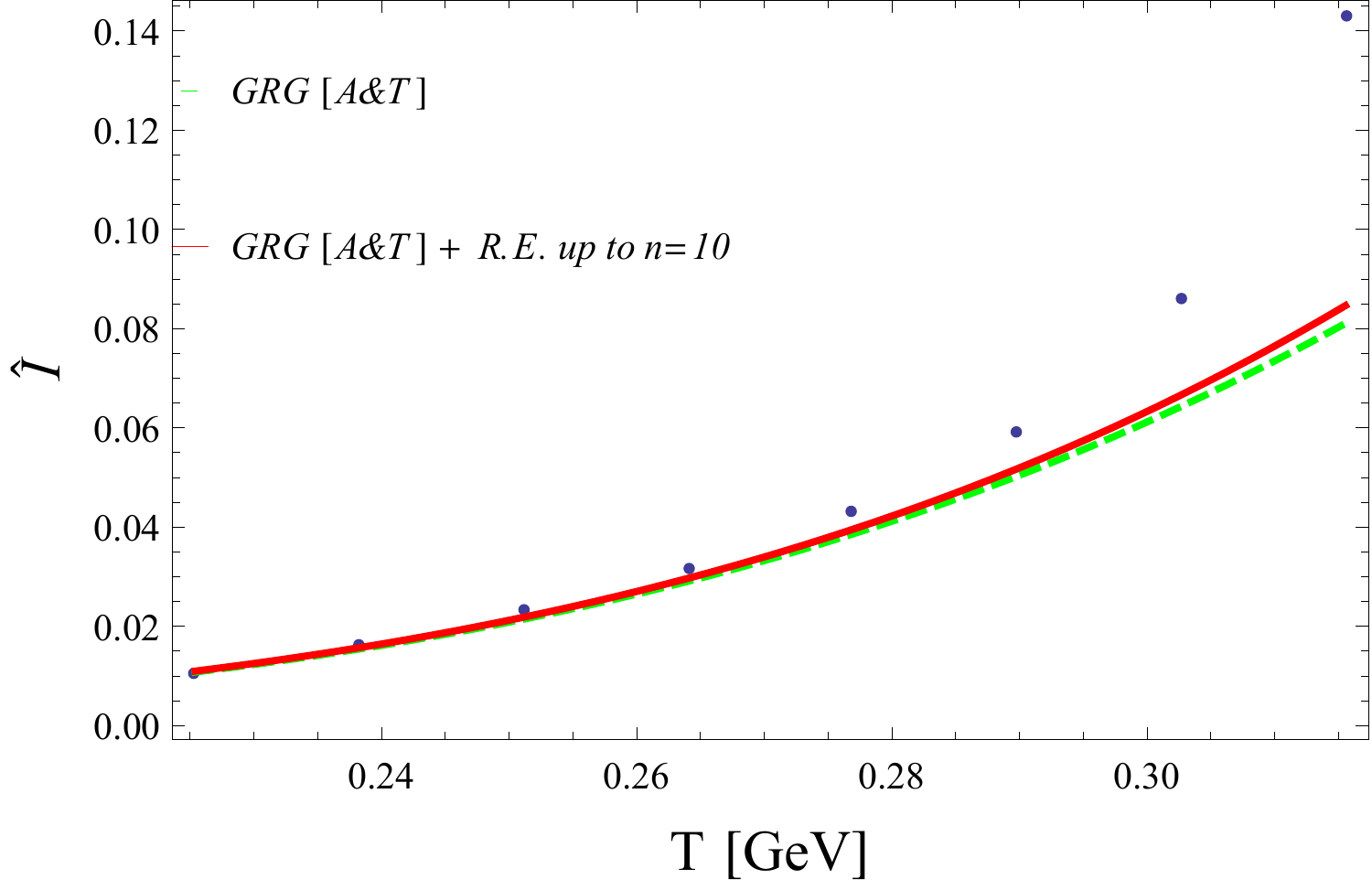}\\
        \caption{Normalized pressure (left) and trace anomaly (right) of the GRG using the masses given by Athenodorou and Teper \cite{Athenodorou:2020ani} (green, dashed) and upon inclusion of the excited states up to $n=10$ for all $J^{PC}$ of that work (red, full). The two plots are compared with the blue dots taken from \cite{Borsanyi:2012ve}. As visible, the increment due to excited states is minimal.}
        \label{Pexcited}
   \end{figure*}

Next, we use these Regge trajectories to evaluate the pressure of the GRG by including glueball masses up to $n=10$ for each quantum numbers listed in Table \ref{tablemass} (that is, not only for those used in the fit of Eq. (\ref{chi2}), but for all listed channels, upon assuming the same slope $a$).   
As it is visible form Fig. \ref{Pexcited}, the contributions of these heavier glueballs to the pressure and trace anomaly turn out to be very small. Their effect becomes visible in the plots only when approaching $T_c$, but also in that region it is quite small. 

It should be also stressed that this approach is not expected to be precise in predicting the masses of higher exited glueballs, but may be regarded as sufficient to estimate the entity of the contribution of such heavy states to the thermodynamic quantities under analysis.


\begin{table}[ht]
    \centering
\begin{tabular}{|c|c|c|c|c|}

\hline
\multicolumn{4}{|c|}{  Glueball spectrum compared to the fit in Eq. (\ref{chi2})}
   & \multirow{2}{*}{ Parameters [$\text{GeV}^2$]}\\
\cline{1-4}
$n\,J^{PC}$&  $m$ [GeV] (from \cite{Athenodorou:2020ani}) & Fit [GeV] &$\chi_{i}^2$
   & \\
\hline
 
    $\textbf{1\, 0}^{++}$\,& $1.653(26)$ &$1.647(25)$&  $0.04$   & \multirow{2}{*}{$b_{0^{++}}=-2.78\pm0.21$}\\
\cline{1-4}
   $\textbf{2\, 0}^{++}$ & $2.842(40)$  &$2.865(30)$& $0.3$& \\
\hline
 
    $\textbf{1\, 2}^{++}$ & $2.376(32)$  &$2.367(30)$ &$0.08$  & \multirow{2}{*}{$b_{2^{++}}=-10.87\pm 0.57$}\\
\cline{1-4}
   $\textbf{2\, 2}^{++}$&  $3.30(5)$ &$3.33(3)$ &$0.38$ & \\
\hline
 
    $\textbf{1\, 0}^{-+}$& $2.561(40)$  &$2.572(38)$ &$0.08$    & \multirow{2}{*}{$b_{0^{-+}}=1.12\pm 0.27$}\\
\cline{1-4}
   $\textbf{2\, 0}^{-+}$ & $3.54(8)$  &$3.48(4)$&$0.57$ & \\
   \hline
 
    $\textbf{1\, 2}^{-+}$ &  $3.07(6)$ &$3.11(5)$ &$0.52$  & \multirow{2}{*}{$b_{2^{-+}}=-6.79\pm0.66$}\\
\cline{1-4}
   $\textbf{2\, 2}^{-+}$ & $3.97(7)$ &$3.90(4)$ & $1.10$  &\\
   \hline
 
    $\textbf{1\, 1}^{+-}$ & $2.944(42)$ &$2.955(37)$  &$0.07$  & \multirow{2}{*}{$b_{1^{+-}}=-2.25\pm0.45$}\\
\cline{1-4}
   $\textbf{2\, 1}^{+-}$&  $3.80(6)$ &$3.77(3)$& $0.23$ & \\

\hline

     && &$\chi_{tot}^2$=3.38& $a=5.49\pm0.17$\\
\hline

\end{tabular}
    \caption{ Fit summary obtained from Eq. (\ref{chi2}). For the glueball masses we assume Eq. (\ref{eq:regge}).}
    \label{tab:fit}
\end{table}

\subsection{Glueball-glueball interactions}
\label{int}
 The S-matrix or phase-space formalism \cite{Dashen:1969ep} has been widely applied in studies of hadrons  at nonzero temperature, e.g. Refs. \cite{Venugopalan:1992hy,Weinhold:1996ts,Weinhold:1997ig,Broniowski:2003ax,Broniowski:2015oha,Lo:2017ldt,Lo:2017sde,Lo:2017lym,Dash:2018can,Dash:2018mep,Lo:2019who,Samanta:2020pez,Samanta:2021vgt} and refs. therein. It allows us to calculate the effect of the interaction on the pressure (as well as other thermodynamic quantities) via the derivative of the scattering phase-shift, since the latter is proportional to the density of states. Thus, both attraction or repulsion as well as the finite width of resonances can be consistently taken into account.
 In the limit in which the interaction is small, the pressure calculated within this formalism reduces to the pressure of a free gas. Thus, this approach offers a justification of the HRG free gas in full QCD and the GRG free gas in YM and offers also a way to go beyond the free gas approximation.

In general, one can study the scattering of two arbitrary glueballs. Here, for definiteness, we consider the glueball-glueball interaction of the two lightest glueballs (scalar-scalar $0^{++}$ and tensor-tensor $2^{++}$ interactions), since they are expected to be the dominant ones.

\begin{itemize}
\item Contribution from the interaction between two scalar glueballs.
\end{itemize}

The scalar-scalar interaction contribution to the the pressure reads
\begin{equation}
    \hat{p}_{0^{++}0^{++}}^{\mathrm{int}} = -\frac{1}{T^3}  \sum_{l=0}^{\infty} \int_{2m_{0^{++}}}^{\infty} dx \dfrac{2l+1}{\pi} \dfrac{d \delta_l^{0^{++}0^{++}} (x)}{d x} \int \frac{d^3k}{(2\pi)^3}\ln{\Big(1-e^{- \beta\frac{\sqrt{k^2+x^2}}{T} }\Big)}
   +\hat{p}_B \text{ ,}
    \label{pint0}
\end{equation}
where the integration is over the square of one of the Mandelstam variables $x=\sqrt{s}$ and $\delta_l^{0^{++}0^{++}}(x)$ is the $l$-th wave phase shift of the the scattering of two scalar glueballs with  $J^{PC}=0^{++}$. Above, the second term $\hat{p}_B$ refers to the pressure of an eventually existing bound state of two scalar glueballs.

Indeed, according to Ref. \cite{Trotti,Trotti:2021nns,Trotti:2022ukp} such a bound state of two scalar glueballs -called glueballonium- may exist when using the dilaton potential to describe the mutual interaction between scalar glueballs \cite{Migdal:1982jp}, see also the appendix \ref{app:TT} for a brief recall. Yet, the effect of this glueballonium state on the thermodynamics is not large. This is due to the fact that the pressure is a continuous function of the interaction strength, thus no sudden jump in the pressure occurs when a bound state forms. The contribution of the jump in the pressure due to the formation of a bound state is partly cancelled by an analogous jump -but with opposite sign- of the phase-shift contribution above the threshold \cite{Ortega:2017hpw,Samanta:2021vgt,Samanta:2020pez}. 

\begin{figure*}[h]
        \centering
       \includegraphics[scale=0.4]{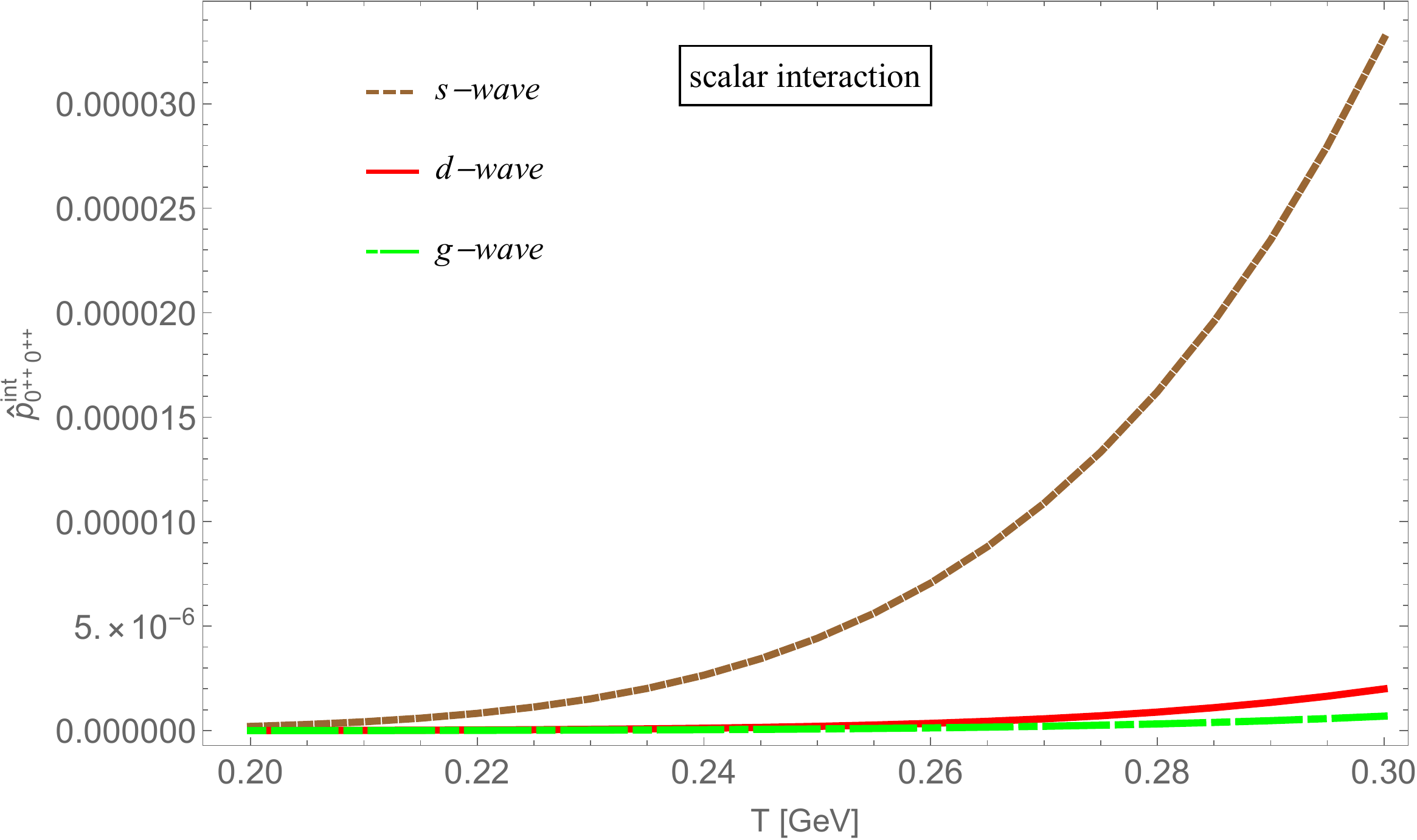}\\
        \caption{Scalar-scalar glueball interaction contribution to the pressure for three different waves: brown (dashed) for s-wave, red (full) for d-wave and green (dotdashed) for g-wave. }
        \label{Pscalar}
   \end{figure*}

Figure \ref{Pscalar} shows the $l=0,2,4$-wave pressure terms due to the interaction between two scalar glueballs obtained by using the phase-shifts evaluated in Ref. \cite{Trotti}. We use $m_{0^{++}}$=1.7 GeV (in agreement with Refs. \cite{Chen:2005,Athenodorou:2020ani}) and $\Lambda_G = 0.4$ GeV, for which a glueballonium with a mass of $3.34$ GeV forms, see details and results in Refs. \cite{Trotti}). 
The $s-$wave ($l=0$) is clearly the dominant one and the contribution of the glueballonium is taken into account. 
When increasing the value of $l$, it  rapidly becomes smaller and smaller, allowing to approximate the total interaction with only the $s-$wave. Although the $l=0$-wave contribution is much larger than the other waves, it is itself negligible when compared to the pressure of the non-interacting $0^{++}$ glueballs.


\begin{figure*}[h]
        \centering
       \includegraphics[scale=0.45]{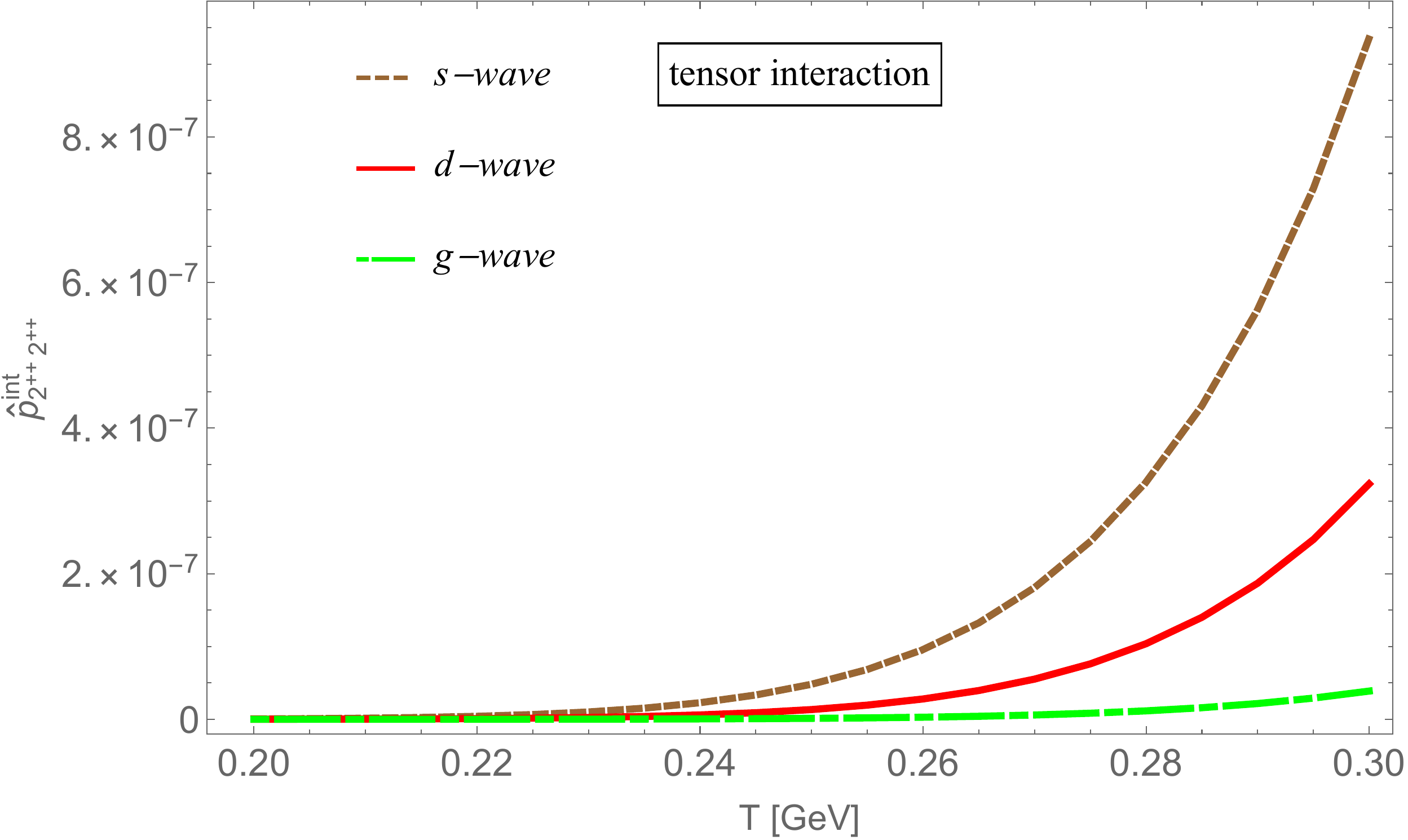} \\
        \caption{Tensor-tensor glueball interaction contribution to the pressure for three different waves:  brown (dashed) for s-wave, red (full) for d-wave and green (dotdashed) for g-wave. Note, each plotted curve is  the sum of the pressure of the same wave for all $J$-values terms with  $J=0,1,2,3,4$. }
        \label{Ptensor}
   \end{figure*}

\begin{itemize}
\item Contribution from the interaction between two tensor glueballs.
\end{itemize}
The next interaction that we take into account addresses two tensor glueballs. The expression takes a form analogous to Eq. (\ref{pint0}):
\begin{equation}
    \hat{p}_{2^{++}2^{++}}^{\text{int}} = -\frac{1}{T^3}\sum_{J=0}^{4}  \sum_{l=0}^{\infty} \int_{2m_2^{++}}^{\infty} dx (2J+1) \dfrac{2l+1}{\pi} \dfrac{d \delta_l^{2^{++}2^{++},J} (x)}{d x} \int \frac{d^3k}{(2\pi)^3}\ln{\Big(1-e^{- \beta\frac{\sqrt{k^2+x^2}}{T} }\Big)}
    \text{ ,}
    \label{pint2}
\end{equation}
where the total spin $J$ of the system ranges between $0$ and $4$ and $m_2^{++} = 2.4$ GeV (in agreement with Refs. \cite{Chen:2005,Athenodorou:2020ani}.)
The phase-shifts $\delta_l^{2^{++}2^{++},J} (x)$ can be calculated by setting an appropriate effective theory for the tensor glueball, see the Appendix \ref{app:TT}. 
As for the scalar case, the s-wave gives a larger contribution to the pressure than the other waves, even though the difference between the two lowest $l-$waves is smaller in the $2^{++}$ case (see Fig. \ref{Ptensor}). Even if the employed effective theory is just the simplest one, the smallness of the result is a solid outcome of the approach. Additional interactions terms among tensor glueballs not included here are not expected to change this result (actually, a four-leg interaction among tensor glueballs, not considered here but expected to be present, would diminish the contribution). 

\begin{itemize}
\item Overall contribution of the interaction.
\end{itemize}


The overall contribution to the pressure of interacting glueballs is obtained by summing up all possible terms:
\begin{equation}
    \hat{p}^{\mathrm{int}}\approx \hat{p}_{0^{++}0^{++}}^{\mathrm{int}}+\hat{p}_{2^{++}2^{++}}^{\mathrm{int}} + ...
    \text{ ,}
    \label{pint}
\end{equation}
where the first two terms involve the two lightest glueballs while dots contain further terms, which are however expected to be smaller because they describe the interactions between more massive glueballs. In fact, the tensor-tensor contribution is already much smaller than the scalar-scalar one. Note, the scattering of two arbitrary glueballs with $J_1$ and $J_2$ can be constructed in a similar way as the tensor-tensor case by considering a total angular momenta $J$ belonging to the interval $[\abs{J_1-J_2},J_1+J_2]$. Moreover, inelasticities and unstable states should be properly taken into account by the interacting terms \cite{Broniowski:2015oha,Lo:2019who,Lo:2021oev,Samanta:2021vgt}.

Summarizing, the overall effect of both the scalar and tensor interaction contributions would be basically invisible in the plots for the pressure and energy density. Thus, the free gas contribution (with the eventual small contribution of heavy glueballs states) seems to provide an accurate approximation for the YM pressure for $T$ below $T_c$.






\section{Conclusion}
\label{concl}
In this work we have studied the GRG model by considering the glueball states observed in lattice simulations. 
In particular, we have shown that the recent lattice mass spectrum of Ref. \cite{Athenodorou:2020ani} delivers a gas which well describes the lattice data of Ref. \cite{Borsanyi:2012ve} below the critical temperature, provided that the latter is consistently chosen to the value $T_c = 323  \pm 18$ MeV.  

Including excited glueballs, which are not yet seen in lattice simulations, via Regge trajectories generates only small differences,  Of course, the effect of such heavy glueballs becomes more relevant for increasing $T$, but their contribution to the pressure turns out to be smaller than 1 percent up to (almost) $T_c$. 

Moreover, we have also computed the contribution to the pressure due the interaction of scalar and tensor glueballs. The effect is even more suppressed than the one of the excited glueballs, thus it can be safely neglected.  This outcome is also in agreement with large-$N_c$ considerations, since glueball-glueball interactions are suppressed in this limit. Namely, the fact that a gas of non-interacting glueballs works pretty well is in agreement with the view that $N_c=3$ can be seen as a `large' number \cite{Panero:2009tv,Lucini:2013qja}. This is not always the case, as recently shown in the large-$N_c$ study of the critical point of the phase diagram in Ref. \cite{Kovacs:2022zcl}.

In conclusion, our results imply that the YM confined phase is relatively simple, since it is dominated by a free gas of glueballs. In turn, the fact that the GRG with the masses of Ref. \cite{Athenodorou:2020ani} works well, can be interpreted as a hint that those masses are consistent with the thermodynamic results.

\section*{Acknowledgement}
We are thankful to Vanamali Shastry, Subhasis Samanta and Markus Huber for useful discussion.  S. J. and E.T. acknowledge
financial support through the project “Development Accelerator of the Jan Kochanowski University of
Kielce”, co-financed by the European Union under the European Social Fund, with no. POWR.03.05.00-00-Z212 / 18.
F. G. acknowledges support from the Polish National Science Centre (NCN)
through the OPUS project no 2019/33/B/ST2/00613. 

\appendix
\section{Scalar and tensor glueball-glueball scattering}
\label{app:TT}
Let us consider the two lightest glueballs, the scalar glueball $G$ and the tensor glueball $G_2 \equiv G_{2}^{\mu\nu}$.
We consider an effective potential for these two fields given by 
\begin{equation}
V_{eff}(G,G_2)= V_{dil}(G)-\frac{\alpha}{2} G^2G_{2,\mu\nu}G_2^{\mu\nu}
\text{ ,}%
\label{eff}
\end{equation}
where the dilaton potential reads \cite{Migdal:1982jp,Schechter:1979}:
\begin{equation}
V_{dil}(G)=\frac{1}{4}\lambda\left(  G^{4}\ln\left\vert
\frac{G}{\Lambda_{G}}\right\vert -\frac{G^{4}}{4}\right)  \text{  .}%
\label{potential}%
\end{equation}
The effective potential contains only one dimensional parameter, the scale $\Lambda_G \sim 0.4$ GeV, which mimics the trace anomaly of YM theory. In addition, the dimensionless parameters $\lambda$ and $\alpha$ are introduced.
As usual, the dilaton/glueball field $G$ develops a nonzero vacuum's expectation value $\Lambda_G$, implying that the shift $G\longrightarrow G+\Lambda_G$ needs to be applied.
As a consequence, the field $G$ gets a mass $m_{0^{++}}^2 =  m_G^2 = \lambda \Lambda_G^2$, while the tensor fied $G_2$ gets a mass $m_{2^{++}}^2 = m_{G_2}^2 = \alpha \Lambda_G^2$. 
Upon using $m_G \sim 1.7$ GeV and $ m_{G_2} \sim 2.4$ GeV, all parameters are fixed.

The  scattering between two scalar glueballs has been extensively described in Ref. \cite{Trotti} and here we just recall its more relevant features (for a recent 
 analogous application to Higgs-Higgs scattering, see Ref. \cite{Shastry:2022goq}).
The tree-level scattering amplitude is based on the three-leg and four-leg interactions emerging when expanding the dilaton potential around its minimum:
\begin{align}
V_{dil}(G)  &=-\frac{1}{16} \Lambda_{G}^4+\frac{1}{2}m_{G}^{2}G^{2}+\frac{1}{3!}\left(
5\frac{m_{G}^{2}}{\Lambda_{G}}\right)  G^{3}+\frac{1}{4!}\left(  11\frac
{m_{G}^{2}}{\Lambda_{G}^{2}}\right)  G^{4}+... \text{ .}
\label{vdil}
\end{align}
In Refs. \cite{Trotti,Trotti:2021nns,Samanta:2021vgt,Shastry:2022goq} two unitarization procedures (the well-known on-shell and $N/D$ ones) are implemented to determine the phase-shift and then, upon using  Eq. (\ref{pint0}), the contributions to the pressure. For the parameter $\Lambda_G \sim 0.4$ GeV, a glueballonium (bound state of two scalar glueballs) exists in the s-wave and is consistently taken into account.   
The connection between the dilaton potential and any $l$-th amplitude for the two scalar glueballs scattering have been described in Refs. \cite{Trotti,Trotti:2021nns}. To summarize, the total (tree-level) amplitude can be extracted from the expanded potential in Eq. (\ref{vdil}):
\begin{equation}
A(s,t,u)=-11\frac{m_{G}^{2}}{\Lambda_{G}^{2}}-\left(  5\frac{m_{G}^{2}%
}{\Lambda_{G}}\right)  ^{2}\frac{1}{s-m_{G}^{2}}-\left(  5\frac{m_{G}^{2}%
}{\Lambda_{G}}\right)  ^{2}\frac{1}{t-m_{G}^{2}}-\left(  5\frac{m_{G}^{2}%
}{\Lambda_{G}}\right)  ^{2}\frac{1}{u-m_{G}^{2}} \text{ .}\label{totampl}%
\end{equation}
Then, rewriting the amplitude as a function of the channel $s$ and the scattering angle $\theta$, the $l$-th amplitude reads:
\begin{equation}
A_{\ell}(s)=\frac{1}{2}\int_{-1}^{1}d\cos\theta A(s,\cos\theta)P_{\ell}(\cos\theta)\, \text{ .} \label{l-ampl}
\end{equation}
This can be directly used to obtain the phase shift $\delta_{\ell}(s)$ as:
\begin{equation}
\delta_{\ell}(s) =\frac{1}{2}\arg\left[  1+2i  \frac{k}{16\pi \sqrt{s}} A_{\ell}(s)\right]  , \label{eq:phasearg}
\end{equation}
where $k$ is the 3-momentum of any particle in the center of mass frame. The results for the phase-shifts (used to obtain the contributions of the pressure) can be found in Ref. \cite{Trotti}. 

As mentioned in the main text, a bound state may form. It turns out that, upon keeping $m_G = 1.7$ GeV fixed, this is the case for $\Lambda_G  \leq 0.504$ GeV. For the critical value  $\Lambda_{G,crit}  = 0.504$ GeV the bound state has a mass of $2m_G$, which decreases upon decreasing $\Lambda_{G}$.  For $\Lambda_{G}  = 0.4$ GeV the bound state has a mass of $m_B = 3.34$ GeV. The results presented in Fig. 5 include this contribution. 
In general, the contribution $p_B$ of the glueballonium bound state with a mass $m_B$ contained in Eq. \ref{pint0} reads:
\begin{equation}
    \hat{p}_B= -\theta(\Lambda_{G,crit}-\Lambda_G) \frac{1}{T^3}  \int \frac{d^3k}{(2\pi)^3}\ln{\Big(1-e^{- \beta\frac{\sqrt{k^2+m_B^2}}{T} }\Big)}
   \text{ .}
\end{equation}
Note, by choosing a slightly larger $\Lambda_G = 0.55$ GeV, no glueballonium exists since the attraction is not strong enough to generate it, but the contribution to the pressure as depicted in Fig. 5 would be basically unchanged, see \cite{Samanta:2020pez,Samanta:2021vgt} .

Next, we concentrate on the scattering of two tensor glueballs, whose formalism is analogous to the one used in Ref. \cite{Petersen:1971ai} in the case of pion-pion scattering. 
The scattering takes the schematic form (for the $s$-channel):
\begin{center}
\begin{centering}
\begin{tikzpicture}
\begin{feynman}
    \vertex (b);
    \vertex [below left=of b] (a) {a};
    \vertex [above left=of b] (f1) {b};
    \vertex [right=of b] (c);
    \vertex [above right=of c] (f2) {c};
    \vertex [below right=of c] (f3) {d};
    \vertex [right=of c] (f4) {$a+b\longrightarrow c+d$,};
    \diagram* {
      (a) -- [plain] (b) -- [plain] (f1),
      (b) -- [dashed, edge label'=G] (c),
      (c) -- [plain] (f2),
      (c) -- [plain] (f3),
    };
  \end{feynman}
  \end{tikzpicture}
\end{centering}
\end{center}
where $a,b,c,d$ refer to the third-component of the spin $J$ of the ingoing and outgoing tensor glueballs. Namely,  four particles can have a different value of the third component $m_J$ of the spin, under the condition that $m^{a}_J +m^{b}_J (\equiv m^{ab}_J) =m^{cd}_J $. Similar diagrams apply in the $t$- and $u$-channels.


The total amplitude is given by:
\begin{equation}
\bra{cd}T\ket{ab}= A\delta^{ab}\delta^{cd}+  B\delta^{ac}\delta^{bd}+ C\delta^{ad}\delta^{bc},
    \label{eq:totamp}
\end{equation}
where $T$ is the transition matrix, and
\begin{equation}\label{ABC}
    A:=\frac{-4(\alpha \Lambda_G)^2}{s-m_{G}^2}\,,\quad B:=\frac{-4(\alpha \Lambda_G)^2}{t-m_{G}^2}\,,\quad C:=\frac{-4(\alpha \Lambda_G)^2}{u-m_{G}^2} \text{ .}
\end{equation}
Upon using Clebsh-Gordan coefficients listed in the PDG \cite{ParticleDataGroup:2022pth}, the initial state reads: 
\begin{equation}
\ket{J^{ab},m^{ab}_J}=\sum_{m^a_J,m^b_J} \bra{J^a\; m^a_J \;J^b\; m^b_J}\ket{J^{ab}\; m^{ab}_J} \ket{J^{a},m^a_J} \otimes \ket{J^{b},m^b_J} \text{ ,}
\label{eq.clebsh}
\end{equation}
where $J^{a}=J^{b}=2$, $J^{ab}\in [0,4]$, $m^{a}_J \in [-2, +2]$, $m^{b}_J \in [-2, +2]$ and $m^{ab}_J \in [-J^{ab}, +J^{ab}]$. Analogously, the same compact form can be written for the outgoing state. \\
Among all the possible combinations, the amplitudes with a nonzero contribution are those $\bra{J^{cd},m^{cd}_J}T\ket{J^{ab},m^{ab}_J}$, for which $J^{cd}=J^{ab}$ and $m^{cd}_J=m^{ab}_J$, implying 25 nonzero amplitudes.
We denote $T^q$ -$q=0,1,2,3,4$ as the amplitude where the total incoming (and outgoing) spin is $J^{q}=J^{cd}=J^{ab}$. For the state $\ket{J,m_J}$ (and analogously for $\bra{J,m_J}$), we employ the basis $\ket{i}$,  $i \equiv (\rom{1},\rom{2},\rom{3},\rom{4},\rom{5})$ as follows:
\begin{align}
    \ket{2,\pm 1}=\sqrt{\frac{1}{2}}\Big(\ket{\rom{1}} \mp \textbf{i} \ket{\rom{2}}\Big)\,\qquad  
    \ket{2, \pm 2}=-\sqrt{\frac{1}{2}}\Big(\ket{\rom{3}} \mp \textbf{i}\ket{\rom{4}}\Big)\,\qquad  
    \ket{2,0}=\ket{\rom{5}} \text{ .}
\end{align}
In this new basis we can write:
\begin{align}
    \ket{i_a} \otimes \ket{i_b}=\ket{i_a}\ket{i_b},
\end{align}
where the value of any $\bra{i_c}\bra{i_d}T\ket{i_a}\ket{i_b}$ is given in term of the parameters $A, \, B \, \text{and} \, C$, given in Eq. (\ref{ABC}).   
Thus, it follows that, using Eq. (\ref{eq:totamp}), any amplitude $T^q(\equiv T^{ab} \equiv T^{cd})$ can be written as:
\begin{align}
    T^q=\sum_{i_a,i_b,i_c,i_d} \rho_{abcd} \bra{i_c}\bra{i_d}T\ket{i_a}\ket{i_b} \,\qquad   i_a,i_b,i_c,i_d \equiv ( \rom{1},\rom{2},\rom{3},\rom{4},\rom{5}),
\end{align}
where $\rho_{abcd}=\bra{2\; m^a_J \;2\; m^b_J}\ket{J^{ab}\; m^{ab}_J}\cdot\bra{2\; m^c_J \;2 \;m^d_J}\ket{J^{cd} \;m^{cd}_J} $ is the product of the Clebsh-Gordan coefficient of the initial and final state.
Using this formalism, together with the definitions in Eq. (\ref{ABC}), we obtain the following amplitudes:
\begin{align*}\boxed{
    T^4=T^2=B+C\,, \qquad T^3=T^1=B-C \,, \qquad T^0=5A+B+C}
    \text{ ,}
\end{align*}
which are used in Eq. (\ref{pint2}) in order to evaluate the contribution to the pressure of the tensor-tensor interaction. As shown in the main text, this contribution is subleading. 


\bibliographystyle{utphys}
\bibliography{main}
\end{document}